\documentclass[twocolumn,showpacs,aps,prb,superscriptaddress]{revtex4}

\usepackage{graphicx}
\usepackage{amsmath}

\begin{document}

\title{Linear temperature dependence of conductivity in the ``insulating'' 
regime of dilute two-dimensional holes in GaAs}

\author{Hwayong Noh} 
\affiliation{Department of Electrical Engineering, Princeton University,
Princeton, New Jersey 08544}

\author{M. P. Lilly}
\affiliation{Sandia National Laboratories, Albuquerque, New Mexico 87185}

\author{D. C. Tsui}
\affiliation{Department of Electrical Engineering, Princeton University,
Princeton, New Jersey 08544}

\author{J. A. Simmons}
\affiliation{Sandia National Laboratories, Albuquerque, New Mexico 87185}

\author{L. N. Pfeiffer}
\author{K. W. West}
\affiliation{Bell Labs, Lucent Technologies, Murray Hill, New Jersey 07974}

\date{\today}

\begin{abstract}
The conductivity of extremely high mobility dilute 
two-dimensional holes in GaAs changes linearly with temperature
in the insulating side of the metal-insulator transition. 
Hopping conduction, characterized by an exponentially decreasing 
conductivity with decreasing temperature, is not observed when the
conductivity is smaller than $e^{2}/h$.  
We suggest that strong interactions in a regime close
to the Wigner crystallization must be playing a role in the unusual 
transport.
\end{abstract}
\maketitle

In recent years, there has been great interest in the transport of
low density two-dimensional (2D) charge carrier systems.
In contradiction to the the scaling theory of localization\cite{loc}
which predicted that all states in 2D in the absence of electron-electron 
interactions are localized,  
experimental observations of ``metallic'' behavior showing increasing 
conductivity ($\sigma$) with decreasing temperature ($T$) ($d\sigma/dT<0$)
and an apparent metal-insulator transition (MIT)
as the carrier density is lowered
have been reported on many low density, low disorder 2D systems.\cite{mit}
Although there have been many experimental and theoretical studies trying 
to understand this metallic behavior, there is still no consensus on its origin
to date.

On the other hand, the ``insulating'' behavior ($d\sigma/dT>0$) for lower
densities in these low disorder systems has not been studied as extensively
as the metallic behavior.
In highly disordered systems, the transport in the insulating regime is 
described by variable range hopping\cite{mott,es} (VRH) among the localized 
states, and the $T$-dependence of $\sigma$ is
expressed as $\sigma(T)=\sigma_{0}exp[-(T_{0}/T)^{x}]$, where $\sigma_{0}$ 
is a prefactor and $T_{0}$ is a characteristic temperature. The exponent $x$
depends on the density of states (DOS) at the Fermi energy. For a constant
DOS in the absence of interactions, $x=1/3$ corresponding to the Mott 
VRH,\cite{mott} and for a DOS which has a Coulomb gap at the Fermi energy 
due to interactions, $x=1/2$ corresponding to the Efros-Shklovski VRH.\cite{es}
In clean systems with low disorder, however, the origin of the 
``insulating'' behavior ($d\sigma/dT>0$) in low density regime, where
the interactions between carriers are strong,
could be significantly different from that described by
simple localization. 
In the ideal case, it was predicted that 
the ground state of the system in the low density limit
is a Wigner crystal,\cite{wigner} which can be pinned by
even a tiny amount of impurities in the system. 

We have recently studied the MIT
of extremely high mobility dilute 2D holes in GaAs and presented
a detailed analysis for the $T$-dependence of $\sigma$
on the ``metallic'' side of the transition.\cite{noh}
The metallic behavior persisted up to $r_{s}=57$,
much higher than 37 predicted for Wigner crystallization
($r_{s}$ is the interaction energy to Fermi energy ratio, given by
$r_{s}=(p\pi)^{-1/2}m^{*}e^{2}/4\pi\hbar^{2}\epsilon$, where $p$ is
the hole density, $m^{*}$ effective mass, and
$\epsilon$ the dielectric constant).
With the persistence of metallic behavior up to such a large $r_{s}$ value,
the question arises regarding the nature of the ``insulating'' behavior
observed in this 2D system for $r_{s}$ reaching 80.
To address this question, we present the $T$-dependence of $\sigma$ of 
the same sample on the ``insulating'' side of the MIT in this report. 
We show that the conductivity exhibits a linear dependence on temperature
and does not follow the VRH behavior, and discuss its implications
with regard to strong interaction effects
close to or in the Wigner crystallization regime.

The sample used in this study is 
a heterojunction insulated-gate field-effect transistor (HIGFET) 
made on a (100) surface of GaAs.\cite{noh,kane}
A metallic gate, separated by an insulator (AlGaAs)
from the semiconducting GaAs, is used to induce the 2D holes at the
interface between the GaAs and AlGaAs. 
The mobility ($\mu$) of our sample measured at $T=65$ mK
reaches $1.8\times10^{6}$ cm$^{2}$/Vs 
for $p=3.2\times10^{10}$ cm$^{-2}$, which is the highest 
achieved for 2D holes in this low density regime.
This high mobility allows us to measure the temperature
dependence of conductivity down to very low densities reaching
$p=1.5\times10^{9}$ cm$^{-2}$, with $r_{s}$ near 80, and the MIT is 
observed at a critical density of $p_{c}=3\times10^{9}$ cm$^{-2}$, i.e.
$r_{s}=57$. The experiment was done in a dilution refrigerator with a
base temperature around 65 mK.
The resistivity of the 2D holes were measured by a low-frequency lock-in
technique with frequency as low as 0.1 Hz to maintain the out-of-phase
signal less than 5 \%. Excitation current of as small as 50 pA was used 
to ensure that the effect of Joule heating is negligible. 
In addition, the drive current was varied at base temperature
to make sure that the linear response was measured.

\begin{figure}[t]
\begin{center}
\includegraphics[width=3in]{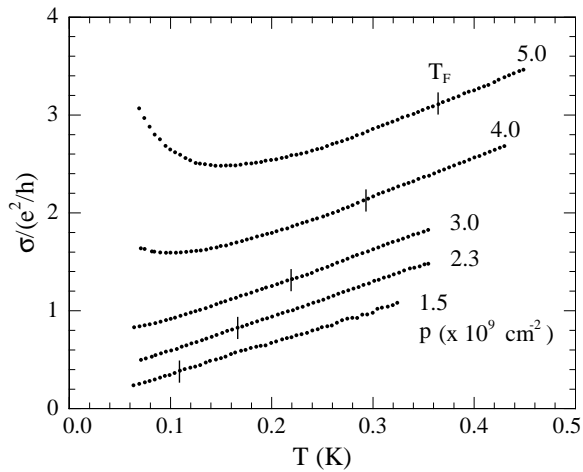}
\end{center}
\caption{\label{1}$\sigma$ vs $T$ 
for $p=$ 5, 4, 3, 2.3, and 1.5$\times 10^{9}$ cm$^{-2}$ from the top.
The Fermi temperature $T_{F}$ is marked by the vertical bar.}
\end{figure}

Figure~\ref{1} shows the $T$-dependence of $\sigma$ for 
five 2D hole densities varying from $5\times 10^{9}$ cm$^{-2}$ ($r_{s}=44$)
to $1.5\times 10^{9}$ cm$^{-2}$ ($r_{s}=80$).
Data for higher densities deep inside the metallic region have been 
presented in Ref~\onlinecite{noh}. The top two traces are on the metallic 
side with $\sigma>e^{2}/h$ in the entire $T$ range. 
For $p=5\times 10^{9}$ cm$^{-2}$,
$\sigma$ decreases with increasing $T$ from 0.065 K to 0.15 K. 
This decrease of $\sigma$ with increasing $T$ ($d\sigma/dT<0$)
is typical of the metallic behavior, the physical origin of which
has been the focus of much recent research on 2D MIT.
For $T>0.15$ K, $\sigma$ increases with increasing $T$ and
the increase becomes linear for $T>0.25$ K.
This linear dependence of $\sigma$ on $T$
continues for $T$ exceeding $T_{F}$, the Fermi temperature of the
2D hole gas (marked by the vertical bar in Fig.~\ref{1}),
which is 0.36 K for this density.
Das Sarma and Hwang\cite{das_sarma} explained this change in $T$-dependence
of $\sigma$ from $d\sigma/dT<0$ to $d\sigma/dT>0$ as a quantum-to-classical
crossover. In the classical regime, when $T\gg T_{F}$, scattering of the
2D carriers by charged impurities would give rise to linear increase of
$\sigma$ with increasing $T$. We note that the linear dependence starts
below $T_{F}$ in our data, and the crossover happens at a temperature
about one third of $T_{F}$.

For $p=4\times 10^{9}$ cm$^{-2}$ ($T_{F}=0.29$ K), 
similar $T$-dependence of $\sigma$
is observed, although the metallic behavior is much weaker, and
the quantum-to-classical crossover occurs at lower $T$ ($\sim0.1$ K)
as expected.
The linear dependence sets in at $T=0.22$ K.
However, for $p=3\times10^{9}$ cm$^{-2}$ ($r_{s}=57$), the third trace
from the top, $d\sigma/dT>0$ is observed in all our $T$ range.
The metallic behavior ($d\sigma/dT<0$) is not seen in the data
down to the lowest $T$, and for $T<0.13$ K $\sigma$ becomes smaller than 
$e^{2}/h$, suggestive of an insulator, although strictly linear dependence 
is not seen for $T<0.1$ K.

The bottom two traces show insulating behavior in that
$\sigma$ is observed to decrease with decreasing $T$, albeit only linearly.
For $p=2.3\times 10^{9}$ cm$^{-2}$ ($r_{s}=65$), $\sigma<e^{2}/h$ for
$T<0.22$ K and the linear dependence extends down to the base temperature.
This linear dependence appears to be the same as
that seen in the high hole density metallic side ($5\times10^{9}$ cm$^{-2}$
and $4\times10^{9}$ cm$^{-2}$), and the same linear dependence is seen
for $\sigma<e^{2}/h$ as well as for $\sigma>e^{2}/h$.
Further reducing the density down to $1.5\times 10^{9}$ cm$^{-2}$ ($r_{s}=80$)
does not change the $T$-dependence very much, but simply decreases the 
magnitude of $\sigma$, making the $T$ range where $\sigma<e^{2}/h$
wider.

\begin{figure}[t]
\begin{center}
\includegraphics[width=3in]{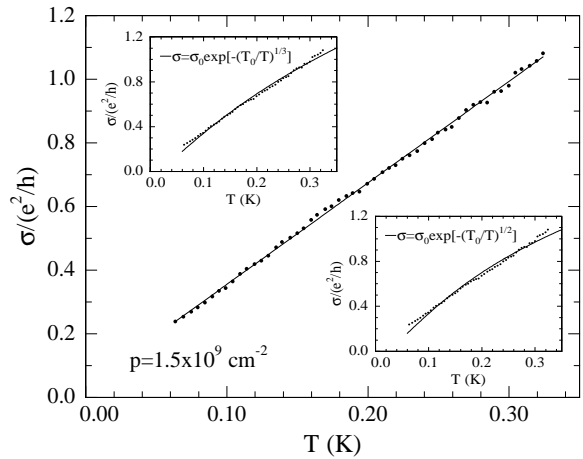}
\end{center}
\caption{\label{2}(a) $\sigma$ vs $T$ for $p=1.5\times10^{9}$ cm$^{-2}$.
Insets: Trial fitting using the VRH model with (a) x=1/3 and (b) x=1/2.}
\end{figure}

In Fig.~\ref{2}, we show the
$T$-dependence of $\sigma$ for $p=1.5\times 10^{9}$ cm$^{-2}$ in more detail.
Clearly, $\sigma$ depends linearly on $T$ (as shown by the solid line)
in the entire $T$ range, and $\sigma<e^{2}/h$ in almost the same range.
When we tried to 
fit the data with the VRH model with exponent $x$ equal to either 1/3 or 1/2,
none of them produced a good fit, as shown in the insets.
This suggests that the transition to an ``insulating'' behavior 
as we lower the 2D hole density in 
our sample, determined from the sign of $d\sigma/dT$, is not 
a disorder induced transition.
While one can argue that further decreasing the density will eventually
lead to an exponentially localized behavior, 
strong interactions (with $r_{s}=80$) which must be 
present for this density make it unlikely that the transport can be 
described by single particle localization.

For a 2D electronic system with
such large $r_{s}$ value, formation of a Wigner crystal should be
considered. When the Wigner crystal is pinned by disorder,
the transport is expected through thermal activation of the 
crystal over the pinning potential. In this case, $\sigma$ should decrease
exponentially with decreasing $T$. Our data, on the contrary, shows a
linear $T$-dependence in the $\sigma<e^{2}/h$ insulating regime.
Here, we should note that reducing the carrier density
not only increases $r_{s}$, but also decreases $T_{F}$, thus decreases
the quantum-to-classical crossover temperature.
In our data, for hole densities $p=5\times 10^{9}$ cm$^{-2}$ and
$p=4\times 10^{9}$ cm$^{-2}$, the linear dependence of
$\sigma$ on $T$ predicted for the $T\gg T_{F}$ classical regime actually 
extends down to a $T$ even below $T_{F}$, suggesting that the
crossover from classical to the degenerate regime seems to occur much
below $T_{F}$. In any case, for a classical 2D electronic system, 
the relevant interaction parameter is $\Gamma$,
which is given by $\Gamma=e^{2}\sqrt{\pi n}/4\pi\epsilon k_{B}T$, with
$n$ being the carrier density.
This parameter is equivalent to $r_{s}$ in the degenerate regime,
with the kinetic energy replaced by $k_{B}T$.
Studies of electrons on liquid Helium\cite{grimes,mehrotra,morf,gann}
have shown that transition to the 
Wigner crystal is around $\Gamma=127$.
In our sample, for the lowest hole density,
$p=1.5\times 10^{9}$ cm$^{-2}$, $T_{F}=0.11$ K.
$\Gamma$ is between 80 and 27 for $T$ from 0.11 K to 0.32 K, 
and equals 146 at $T=0.06$ K.
Therefore, Wigner crystallization is not likely to occur in most
of our $T$ range until $T$ reaches close to the base temperature.
However, a more recent electron-on-helium experiment\cite{djerfi} has reported
residual signature of the Wigner crystal up to a much higher temperature 
with $\Gamma$ down to 46. Thus, we should expect the 2D holes in our 
sample to be strongly correlated and the Wigner crystallization
physics be important.
The $T$-dependence arising from scattering of the 2D hole
gas by charged impurities, as considered by Das Sarma and 
Hwang, even though remarkably similar to that observed in our experiment,
cannot be the adequate explanation for the insulating regime.

We should also mention that localization corrections to the conductivity 
in the weakly interacting classical regime do not explain 
our data.
Experimental study on weak localization in the classical regime
has been performed for electrons on solid hydrogen surface.\cite{adams}
In this case, the correction to the conductivity is proportional to
$T_{F}/T$, and our data are not consistent with this expectation.
The interaction correction
is proportional to $(T_{F}/T)^{2}$. This term is negligible
compared with that from the weak localization, and
does not explain our data either.

Finally, the linear dependence of $\sigma$ on $T$
observed in the insulating regime of our data
must result from strong interaction effects. 
In this respect, we think that the recent theory by Spivak\cite{spivak} 
may be relevant to our data.
In his theory, the metallic state is a state in which a small
concentration of Wigner crystal droplets are embedded in the Fermi liquid,
and the insulating state corresponds to one in which a small fraction of
Fermi liquid is present in a mostly crystalline phase. 
In the metallic side, 
the resistivity increases linearly in $T$ in the low temperature
degenerate regime as the size of the Wigner crystal droplets grows.
In the insulating side, he argues that the crystal
forms a supersolid, which is not pinned by disorder, and 
$\sigma$ remains finite as $T$ goes to zero.
At high temperatures, Wigner crystal droplets melt and the system
behaves classically. In this case, the resistivity
is inversely proportional to $T$ and $\sigma$ is linear in $T$.
We note that this predicted $T$-dependence
is identical to that by Das Sarma and Hwang, but it is based on strong
correlation effects close to the Wigner crystallization.
The prediction that $\sigma$ remains finite at $T$=0 
in the insulating side is
also significant, since in our data the linear dependence appears to
continue below $T_{F}$ with finite offset.

In summary, we have studied the temperature dependence of the conductivity
in the insulating side of the MIT for extremely high mobility dilute
2D hole system with $r_{s}$ reaching 80. A linear dependence of conductivity
on temperature has been observed in the insulating side. We suggest that
strong interaction effects in both degenerate and classical regime
close to the Wigner crystallization should be considered
to provide an understanding for this unusual transport.

This work has been supported by the NSF and the DOE at Princeton University.

\end{document}